\documentclass[journal, varwidth, notab]{IEEEtran}
\IEEEoverridecommandlockouts

\usepackage{lipsum}

 \usepackage[pdftex]{graphicx}
 \usepackage[font=small,labelfont=bf,center]{caption}

\DeclareCaptionLabelSeparator{period-newline}{\par}
\usepackage{stfloats} %
\usepackage[percent]{overpic}

\usepackage{array}
\usepackage{makecell}

\usepackage{listings}

\usepackage[document]{ragged2e}
\usepackage{ragged2e}
\justifying
\usepackage{enumerate}

\usepackage{amsmath}
\usepackage{siunitx} 
\DeclareSIUnit{\belmilliwatt}{Bm}
\DeclareSIUnit{\dBm}{\deci\belmilliwatt}
\usepackage{nicefrac}
\usepackage{nccmath}

\usepackage{url}

\usepackage[colorlinks=true,linkcolor=blue]{hyperref}	%
\hypersetup{ %
    linkcolor=,
    anchorcolor=,
    citecolor=,
    filecolor=,
    menucolor=,
    runcolor=,
    urlcolor=,
}

\usepackage[utf8]{inputenc}

\usepackage[acronym, toc,numberedsection = nolabel, section = part]{glossaries}
\makenoidxglossaries
\newacronym{adc}{ADC}{Analog to Digital Converter}

\glsdisablehyper

\usepackage{csquotes}
\usepackage[style=ieee,backend=biber,sorting=none]{biblatex}
\DeclareFieldFormat[article,inbook,incollection,inproceedings,patent,thesis,unpublished]{title}{#1}

\let\oldcite\cite
\renewcommand*{\cite}[1]{~\oldcite{#1}}

\newcommand{\comment}[1]{}

\usepackage{titlesec}
\titlespacing\section{0pt}{5pt plus 1pt minus 1pt}{0pt plus 2pt minus 2pt}

\newcommand{\secref}[1]{\hyperref[{#1}]{Section \ref*{#1}}}

\usepackage{amsmath}
\usepackage{cleveref}

\begin{document}

\title{HEEPidermis: a versatile SoC for BioZ recording}

\author{\IEEEauthorblockN{Juan Sapriza$^1$, Beatrice Grassano$^1$, Alessio Naclerio$^2$, Filippo Quadri$^1$, Tommaso Terzano$^1$, David Mallas\'en$^1$, Davide Schiavone$^1$ Robin Leplae$^3$,  Jérémie Moullet$^3$, Alexandre Levisse$^3$, Christoph M\"uller$^3$, Mariagrazia Graziano$^2$, Mat\'ias Miguez$^4$, David Atienza$^1$}

  \IEEEauthorblockA{
  \small
  \begin{tabular}{c}
  $^1$ Embedded Systems Laboratory (ESL), EPFL, Switzerland \\
  $^2$Department of Electronics and Telecommunications, Politecnico di Torino \\
  $^3$ EPFL, Switzerland \\
  $^4$ Engineering Department, Universidad Cat\'olica del Uruguay, Uruguay\\
  Email: juan.sapriza@epfl.ch
  \end{tabular}
  }
\thanks{This work was supported in part by the the Swiss NSF Edge-Companions project (GA No. 10002812), and in part by the Swiss State Secretariat for Education, Research, and Innovation (SERI) through the SwissChips Research Project. This research was partially conducted by ACCESS – AI Chip Center for Emerging Smart Systems, supported by the InnoHK initiative of the Innovation and Technology Commission of the Hong Kong Special Administrative Region Government.
\vskip 0.1in
}
}
\maketitle
\justifying

\begin{abstract}

Biological impedance (BioZ) is an information-packed modality that allows for non-invasive monitoring of health and emotional state. Currently, most research involving tissue impedance is based on bulky or fixed-purpose hardware, which limits the scope of research and the possibilities of experiments. In this work, we present HEEPidermis: a System-on-Chip (SoC) which integrates all the blocks needed for tissue impedance measurement, including two 8-bit, arbitrary-signal current DACs, two VCO-based ADCs, and a RISC-V CPU to enable on-chip feature extraction for closed-loop operation. An event-based sub-sampler improves storage and energy efficiency for long-term recording. In addition to the versatile SoC, the digital back-end and behavioral models of the analog front-end are open-source, allowing fast system-level simulations or repurposing. The SoC was taped out on TSMC 65 nm LP process. 

\end{abstract}

\glsunsetall
\begin{IEEEkeywords}
Bioimpdance, SoC, Open-Source, Edge AI, Biomedical Systems 
\end{IEEEkeywords}
\glsresetall

\section{Introduction} \raggedbottom

\IEEEPARstart{B}{iological} impedance (BioZ) provides a rich source of physiological information, enabling monitoring of properties such as hydration, tissue health, and cell activity\cite{boucsein2012electrodermal}. 
For instance, measuring Electrical Impedance Spectroscopy (EIS) can allow to detect skin-health issues~\cite{rinaldi2021electrical} or crop's needs~\cite{garlando2022ask}; while using a DC current to measure Galvanic Skin Response (GSR) can be used to track emotional distress~\cite{boucsein2012electrodermal}. 

\begin{figure}[t!] 
\centering\includegraphics[width=0.8\linewidth]{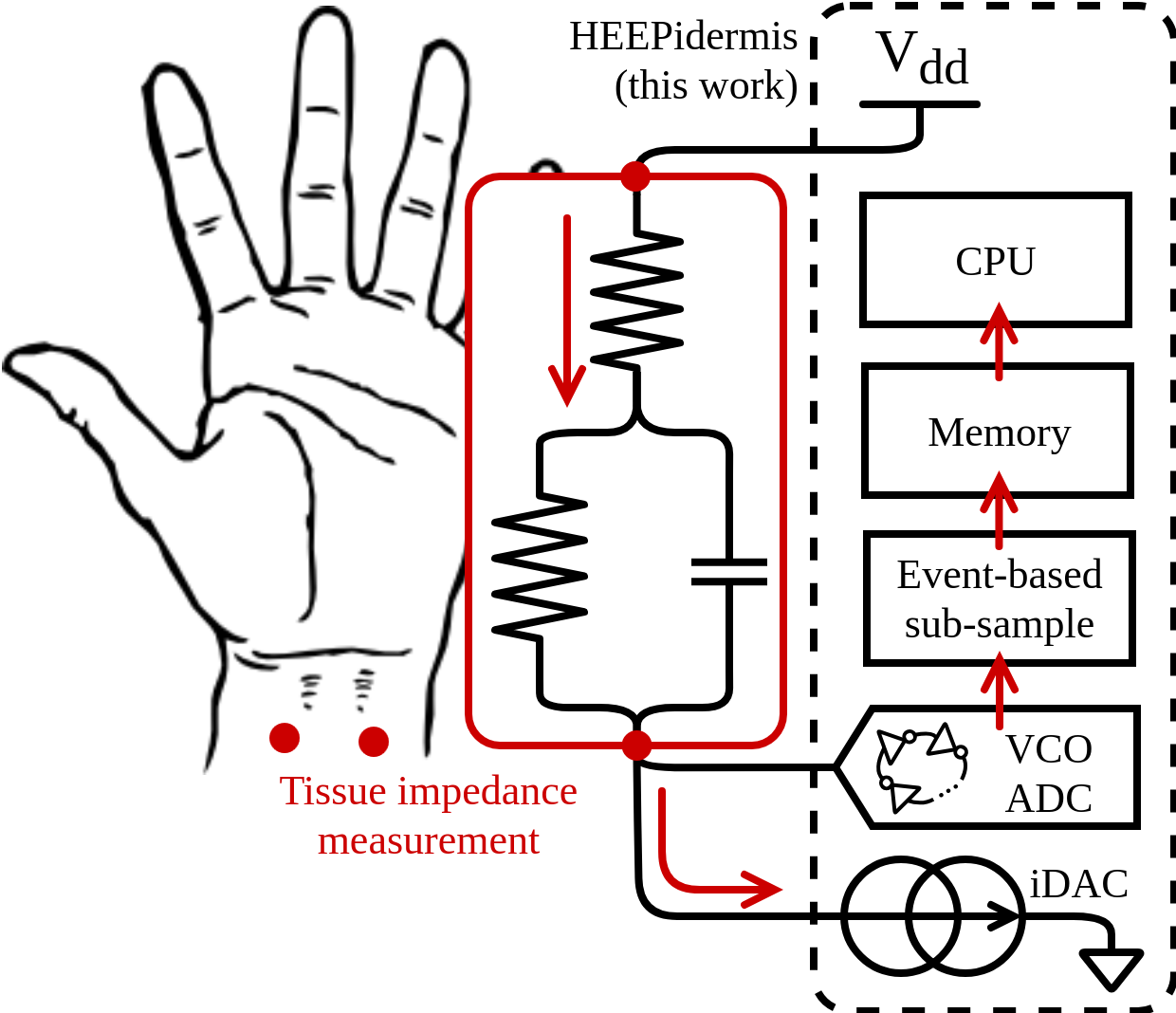}
\caption{BioZ measurement using HEEPidermis: a programmable 8-bit current source drains current from tissue, a VCO-based ADC measures voltage drop. Input data is subjected to an event-based comparator to wake up the CPU for processing or adjusting the current to trade off distortion/power.}
\label{fig:gral-concept}
\end{figure}

A typical impedance measurement, exemplified in \autoref{fig:gral-concept}, requires the application of a known stimulus, e.g., a stable DC current or voltage for resistance measurements or a sinusoidal signal to capture both resistance and reactance through phase analysis, followed by sensing the resulting voltage across the tissue. Skin potential can also be measured from a reference DC voltage (without injection of current) to assess mental engagement~\cite{aminosharieh2023drivers, affanni2020wireless}. 
All these modalities, although similar, vary greatly in dynamic range, required sensitivity, and bandwidth, so state-of-the-art designs are usually tailored to a fixed set of applications~\cite{empatica2025e4,shimmer2025gsr3}. This results in a limited selection of DC current levels and restricted options for AC shape, amplitude, and frequency, which constrains their versatility and adaptability to different experimental requirements. Furthermore, given their fixed-task nature, these devices are rarely ever programmable, restricting the possibility for self-adjustment for long-term recording, or feature extraction for closed-loop therapies. 
This restricts the practicality of long-term, continuous monitoring, which is crucial for capturing the slow, subtle changes in impedance that reflect the underlying biological processes. 

To counter these limitations, in this work we present \mbox{HEEPidermis}\footnote{``HEEPidermis" is a play on words between X-HEEP: the platform it is based on, and ``epidermis": the surface of the skin, where BioZ can be measured.} (\autoref{fig:gral-concept}), a System-on-Chip (SoC) that integrates all necessary functional blocks to perform BioZ recordings and perform on-chip processing of the acquired data. HEEPidermis is based on the open-source platform X\nobreakdashes-HEEP~\cite{machetti2024xheep}, which takes care of the digital back-end for data management. A custom mixed-signal front-end is tightly coupled into the system bus so CPU, DMA or peripherals with bus access can directly command current injection or read ADC samples. 
A behavioral model of HEEPidermis is available\footnote{github.com/esl-epfl/HEEPidermis} and can be tested on RTL simulators supporting non-synthesizable modules. Due to the subsystem-oriented design inherited from the X-HEEP platform, peripherals can be easily modified or replaced to obtain application-specific variants of the SoC. Therefore, the contributions of this work are two-fold: 1) a versatile SoC for impedance measurement, and 2) a modular, open-source platform that can be easily readapted for different low-power recording applications.
HEEPidermis was taped out on the TSMC 65nm LP HVT process.

\section{System overview}

\begin{figure*}[t]
  \centering

  \begin{minipage}[t]{0.72\textwidth}\vspace{0pt}
    \centering
    \begin{overpic}[width=\textwidth]{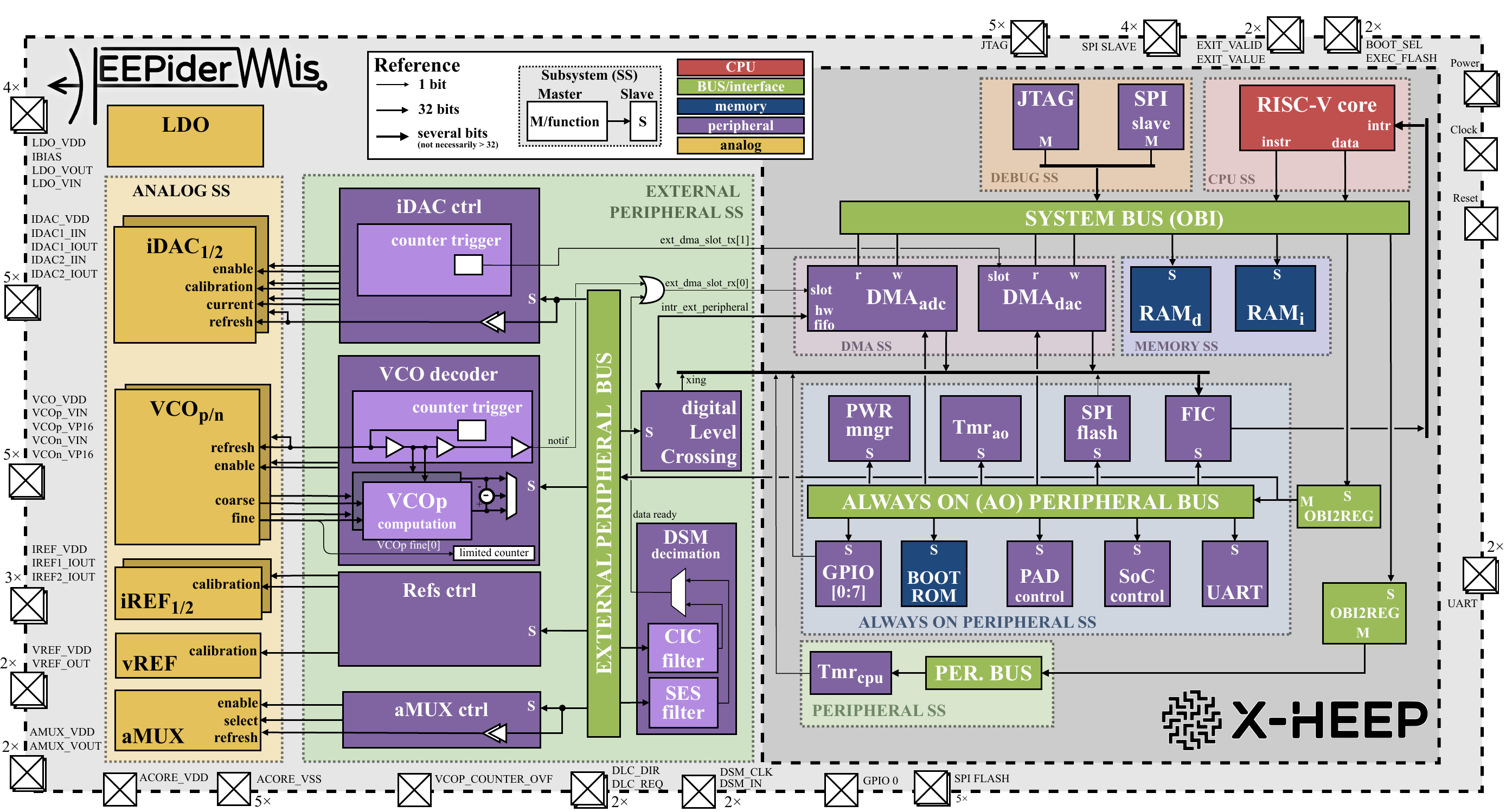}
      \put(-2,50){\small (a)}
    \end{overpic}
  \end{minipage}\hfill
  \begin{minipage}[t]{0.28\textwidth}\vspace{0pt}
    \centering
    \begin{overpic}[width=0.8\linewidth,height=.2\textheight,keepaspectratio]{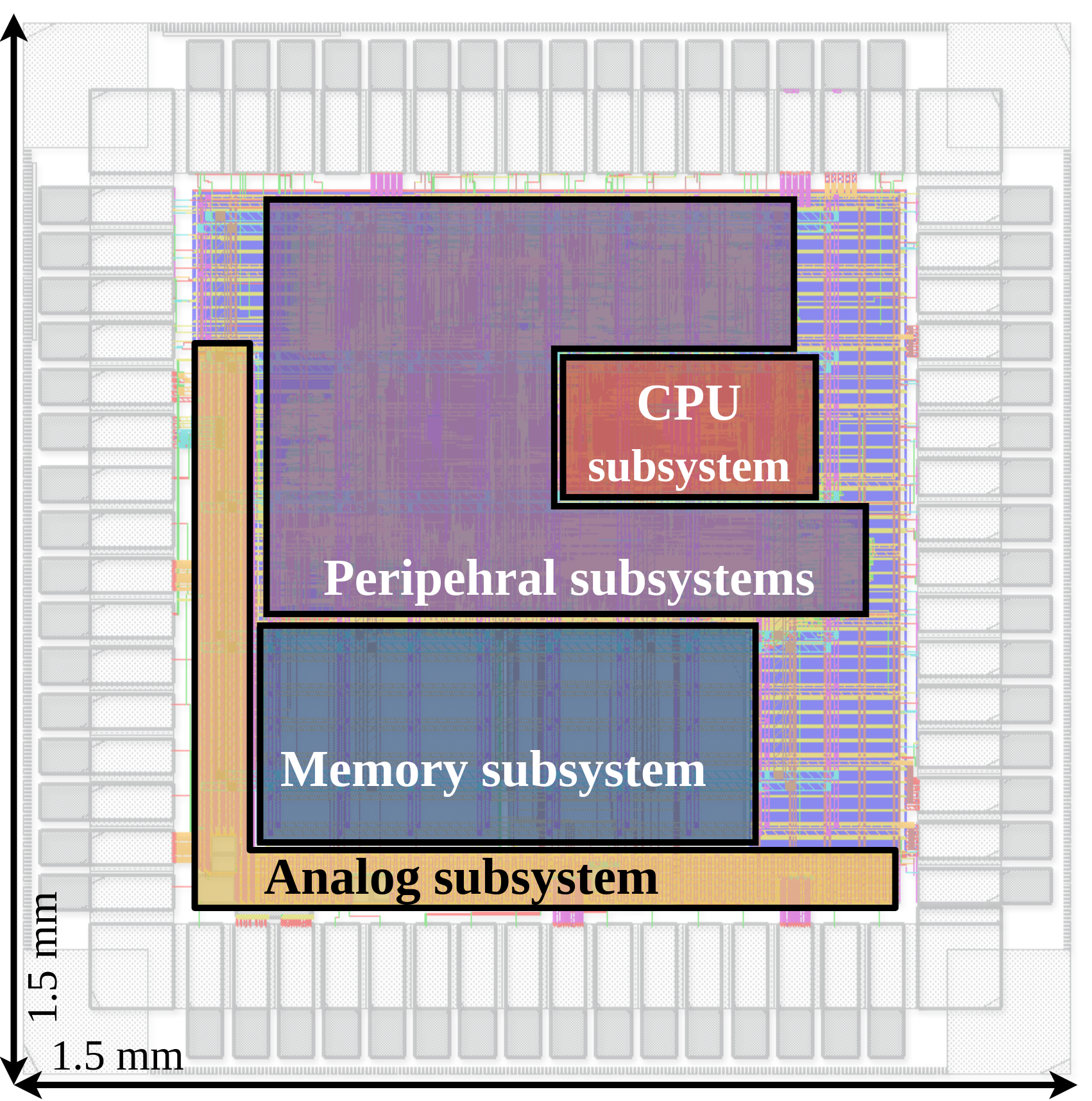}
      \put(-10,90){\small (b)}
    \end{overpic}

    \vspace{0.7em}

    \begin{overpic}[width=\linewidth,height=.50\textheight,keepaspectratio]{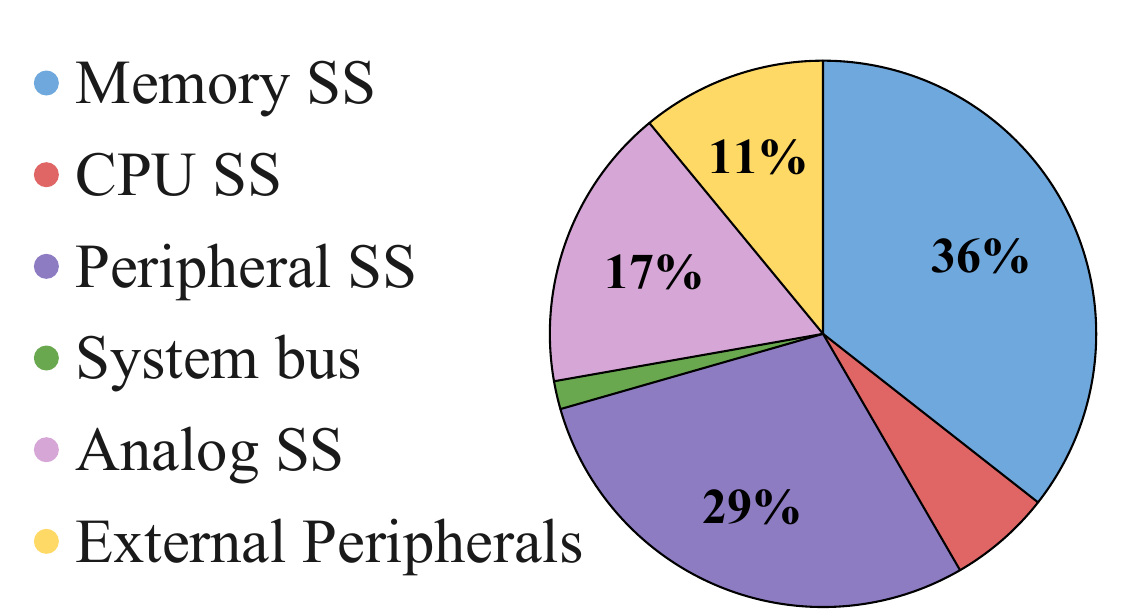}
      \put(1,52){\small (c)}
    \end{overpic}
  \end{minipage}

  \caption{(a) System-level overview of the HEEPidermis SoC. The digital back-end is based on the X-HEEP platform. The analog front-end is controlled through memory-mapped registers (in the external peripherals subsystem. (b) Layout of the fabricated chip with overlayed subsystems. (c) Area breakdown among subsystems.} 
  \label{fig:system-level}
\end{figure*}

\subsection{Digital Back-end}

The digital back-end is based on the X-HEEP platform: an open-source MCU template that enables a modular and easily extensible design,  highlighted on the right, dark-gray half of \autoref{fig:system-level}-a~\cite{machetti2024xheep}. It is configured to use a low-power \texttt{cve32e20} CPU~\cite{cve2} as the main controller, a fully-connected system bus, two separate 16~kiB SRAM banks that allow for parallel instruction and data access, several peripherals as JTAG and SPI to allow a host system to program and read memories (acting as an enhanced ADC), as well as another SPI to allow HEEPidermis to load instructions from an external flash or control external peripherals. Two independent DMA blocks~\cite{terzano2025just} allow to move data between memory and the CPU or the analog front-end. One of the DMAs is coupled to a digital Level-Crossing (dLC) block~\cite{martinez2025reconfigurable}, which acts as a sub-sampler for data coming from the ADCs. It can discard and re-format samples, or monitor the amplitude of the incoming data to wake up the CPU in case of events, enabling an event-based operation of the SoC. 
On the sensing side, HEEPidermis can acquire data through different means: 
\begin{enumerate}
    \item From either of the two on-chip ADCs.
    \item From both on-chip ADCs, in pseudo-differential mode.
    \item From external ADCs through SPI.
    \item From $\Delta\Sigma$ ADCs, which can be decimated with two distinct filters~\cite{martinez2025reconfigurable}. 
\end{enumerate}
All options can be operated autonomously (without CPU intervention) using the DMA, and can be passed through the dLC for sub-sampling. 

\subsection{Analog Front-end}

All analog blocks (with the exception of the LDO) are controlled by memory-mapped registers. The analog subsystem (in yellow in \autoref{fig:system-level}), accounts for less than 17\% of the core area. 

\textbf{LDO} A fixed \SI{0.8}{\volt} LDO is used to supply the current injection. It can also be used to supply the digital core with a lower voltage to reduce power consumption at the expense of insertion of high-frequency noise into the reference. 

\textbf{Voltage reference (vREF)} A \SI{0.8}{\volt} voltage reference for the LDO is integrated. It can alternatively be used as a potential reference for skin potential recording. It can be calibrated through the register interface to a 1\% error. 

\textbf{Current reference (iREF)} A \SI{400}{\nano\ampere} current reference is used to feed the current mirror used in the iDAC. It can be calibrated through the register interface to a 1\% error. 

\begin{figure}[t!]
  \centering

  \begin{minipage}{0.9\linewidth}
    \begin{overpic}[width=\linewidth]{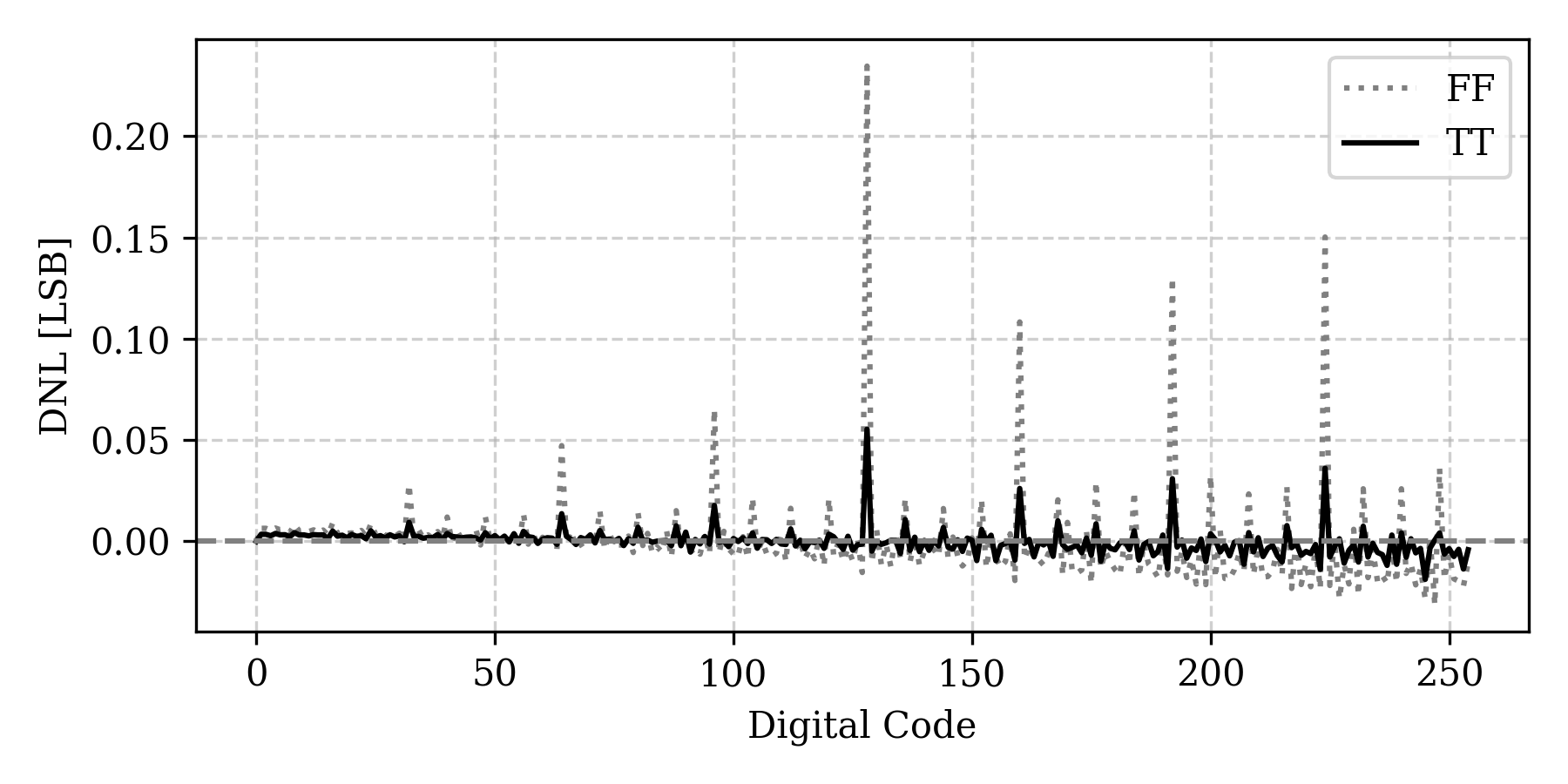}
      \put(1,47){\small (a)} %
    \end{overpic}
  \end{minipage}

  \vspace{-2.3em} %

  \begin{minipage}{0.9\linewidth}
    \begin{overpic}[width=\linewidth]{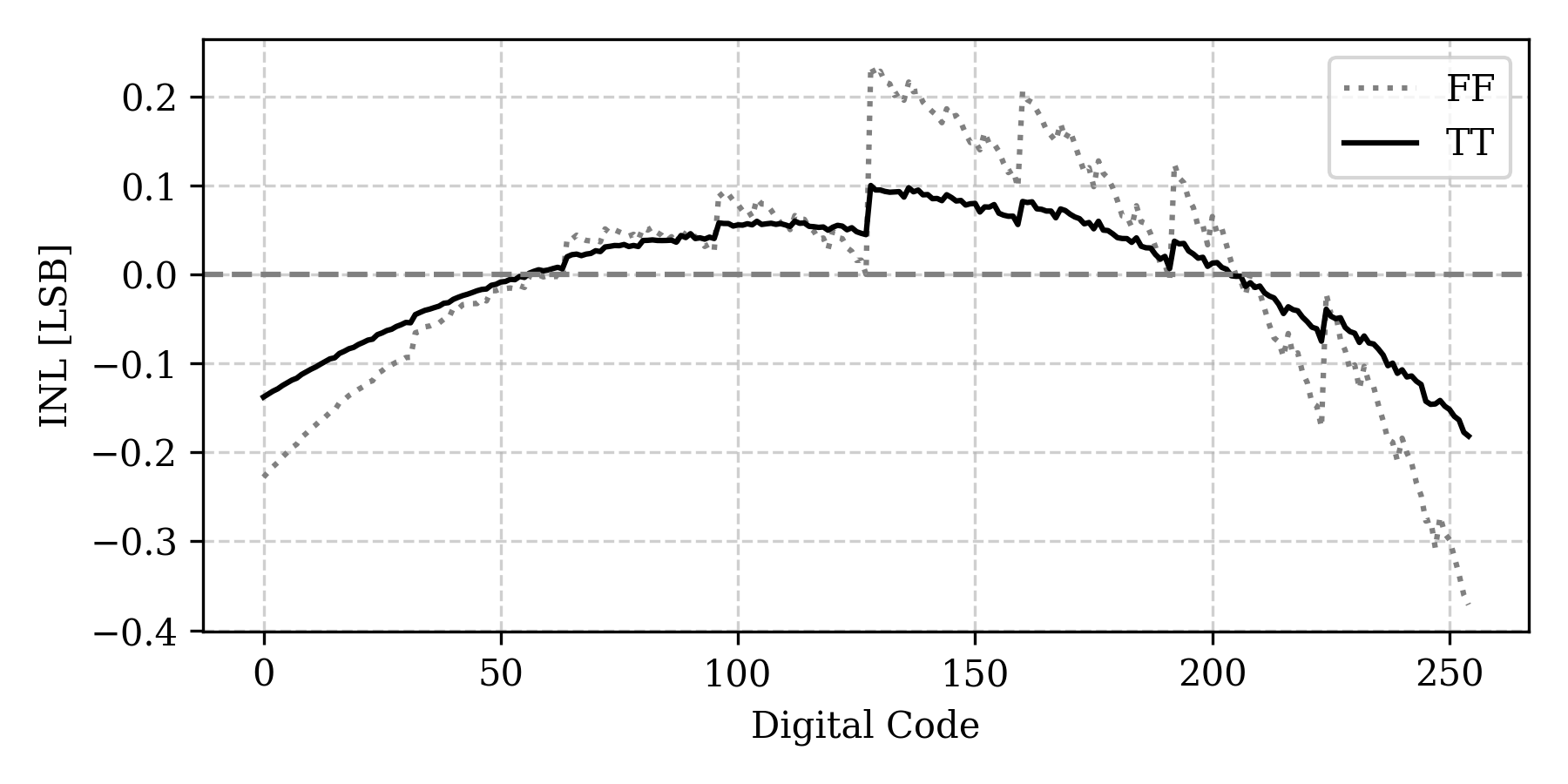}
      \put(1,45){\small (b)} %
    \end{overpic}
  \end{minipage}

  \caption{(a), (b) DNL and INL, respectively, of the iDAC at TT/\SI{20}{\celsius} and FF/\SI{50}{\celsius} (worse) corner cases.}
  \label{fig:dnl_inl}
\end{figure}

\textbf{Current DAC (iDAC)} A pair of identical 8-bit iDACs allow to sink current from the tissue and into ground, from 0 to \SI{10}{\micro\ampere} in steps of \SI{40}{\nano\ampere}. They are implemented as binary current mirrors where the reference branch can be calibrated through the register interface to a 1\% error. Both iDACs are updated simultaneously from a memory-mapped register. The update can be periodic through an integrated timer to obtain arbitrary signal shapes, or on demand from the CPU for stimulation pulses or DC values. The DMA can be used to periodically provide new samples at a rate up to \SI{200}{\kilo\hertz} (for a full-range sweep, limited by the iDACs worst settle time). This allows, for example, the injection of a full-range sine of \SI{1}{\kilo\hertz} with $SQNR=72~\text{dB}$ or \SI{10}{\kilo\hertz} with $SQNR=65~\text{dB}$. As seen in \autoref{fig:dnl_inl}, post-layout simulations show a DNL below 0.25 LSB, and an INL below 0.4 LSB in the worst corner case (FF).  

\begin{figure}[t!]
    \centering
    \includegraphics[width=0.9\linewidth]{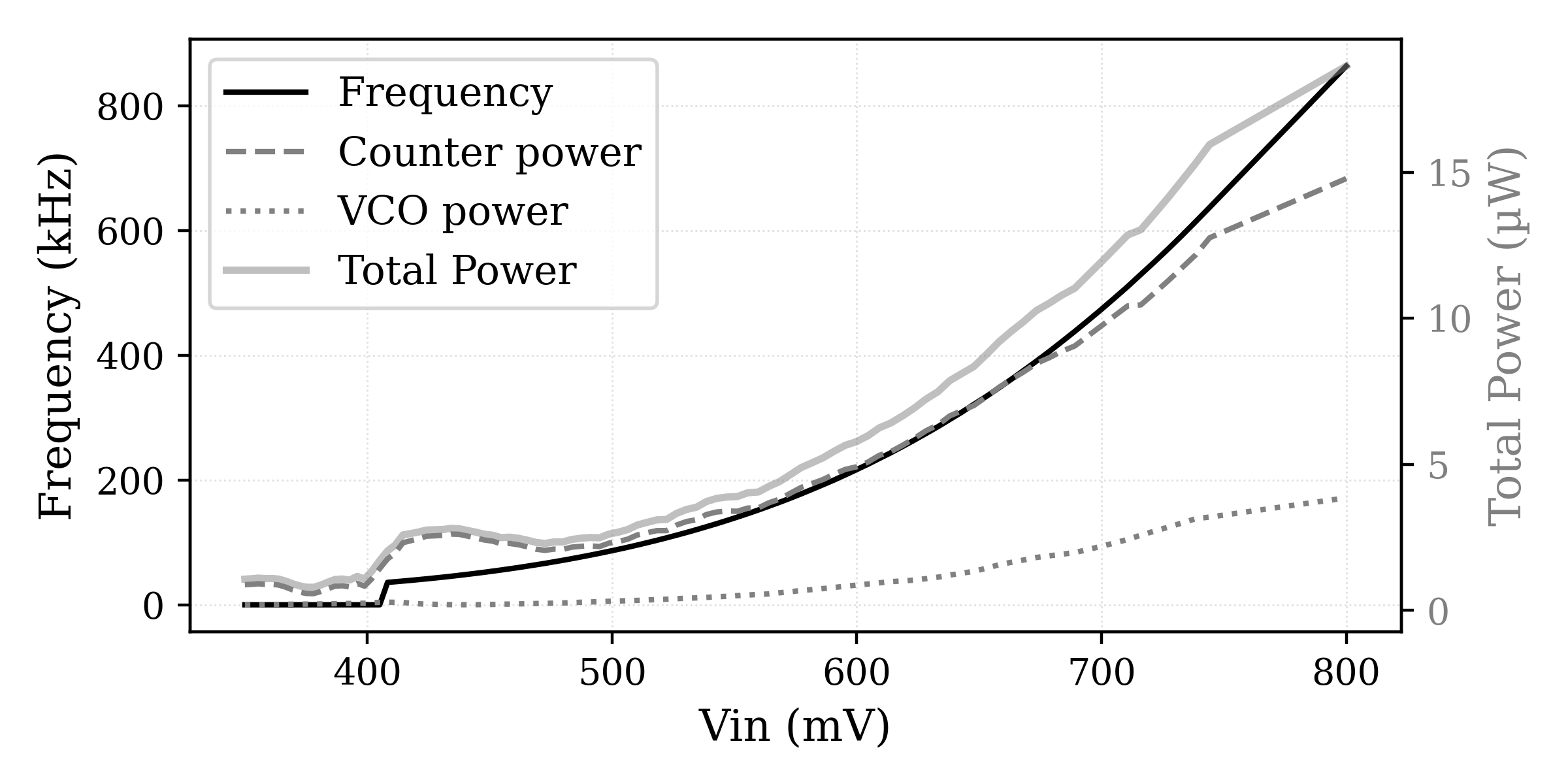}
    \caption{VCO-ADC frequency (left) and ADC power (right). The latter is broke down into the oscillator's power (VCO) and the 26 bit counter after it.}
    \label{fig:vco_transfers}
\end{figure}

\textbf{VCO-based ADC} A pair of Voltage-Controlled Oscillator (VCO)-based ADCs were integrated~\cite{alvero2026vcocare}. They can be used one at a time or in pseudo-differential mode. As seen in \autoref{fig:vco_transfers}, the VCO has an oscillation range from 36-\SI{887}{\kilo\hertz} in the range 408-\SI{800}{\milli\volt}. The output of one of the 31-taps of the oscillator increases a non-reset 26-bit counter that can be sampled either periodically through an integrated timer or on-demand from the CPU. The counter reading is differentiated (and subtracted, in pseudo-differential mode) before being placed into the memory-mapped output register. The outputs of each inverter of the oscillator are also exposed and can be used to compute the phase difference and gain up to 6-bit resolution in the high-performance operational region of the VCO (at low oscillation frequencies this is obscured by phase noise). The counter width was fixed at 26 bits, so the result after adding these 6-bits could fit in a 32-bit word. The large counter size enables longer sampling periods, which improve sensitivity, but does not translate into larger ENOB. During impedance measurement, adjustment of the injected current DC value sets the operating point of the VCO. Higher input voltages will yield a higher gain and thus SNR, but will also drastically increase power up to \SI{20}{\micro\watt}, mainly due to the counter update frequency, as seen in \autoref{fig:vco_transfers}. Power can be reduced by an order of magnitude by lowering the operational point, and SQNR can be compensated for by increasing the oversampling ratio.     

\section{Impedance measurement process}

 The joint use of iDAC, VCO-based ADC and event-based subsampling allows HEEPidermis to be reconfigured to trade-off sensitivity, range, data rate and power, even in real-time, to make up for long-term variations. Two exemplary uses cases are presented to illustrate the SoC's versatility. 

\subsection{Use case 1: GSR measurement}
By injecting a constant DC current, variations in skin conductance can be recorded. This modality has a wide inter-patient dynamic range (level), from \SI{1}{\micro\siemens} to \SI{100}{\micro\siemens}, but changes occur over large periods of time (minutes), and it is atypical that a single patient will present such a large variation. Phasic peaks (spontaneous electrodermal responses) may require a sensitivity of \SI{1}{\nano\siemens} for their shape to be accurately captured during the lowest level periods~\cite{boucsein2012electrodermal}. To avoid distortion introduced from the VCO's non-linearity, the operational point (DC value at the ADC's input) can be chosen by adjusting the injected current according to the measured input voltage. A single VCO can be used and its input can be passed through the dLC. The dLC is configured to compare the input with the desired input range. When the input signal is out of range, the dLC triggers an interrupt that wakes up the CPU, which can adjust the current injected by the iDAC accordingly. 
In this configuration, the digital back-end is mostly quiescent but for sporadic DMA transfers into RAM. 

\autoref{tab:case1} summarizes the power breakdown for GSR (left) at a maximum-range and maximum-sensitivity operational point, where the system achieves a $10\,\text{pS}-100\,\text{nS}$ sensitivity in the range $10\,\text{nS}-100\,\mu\text{S}$, using a \SI{280}{\nano\ampere} current, and sampling twice per second (leaving the digital back-end to go to sleep).   

\begin{table}[b!]
\caption{Power consumption for BioZ recording}
\centering
\setlength{\tabcolsep}{8pt}
\renewcommand{\arraystretch}{1}
\begin{tabular}{l c c}
\hline
\textbf{Component} & \textbf{GSR meas.} ($\mu$W) & \textbf{Impedance meas.} ($\mu W$)\\
\hline
iDACs              & 1 & $2\times 3$ \\
VCOs               & 17  & $2\times 10$   \\
Digital back-end  & 46 & 46  \\
\hline
\textbf{Total}              & \textbf{64} & \textbf{72}
\\\hline
\end{tabular}
\label{tab:case1}
\end{table}

\subsection{Use case 2: Impedance measurement}
To measure impedance, a sine can be injected and its amplitude and phase difference can be measured. Because there is no way to measure the phase of the input signal with respect to the injected one (the ADC is not synchronized with the iDAC), a cancellation trick can be used.
The output of one of the iDACs sinks current from the tissue while the other does so through a reference resistor. Each output is also connected to one of the two VCO-ADCs, which are used simultaneously in pseudo-differential mode. By making two measurements with two different known phase differences between the injected sines, the amplitude and phase variation through the tissue can be inferred. %

\begin{figure}[t!] 
\centering\includegraphics[width=0.9\linewidth]{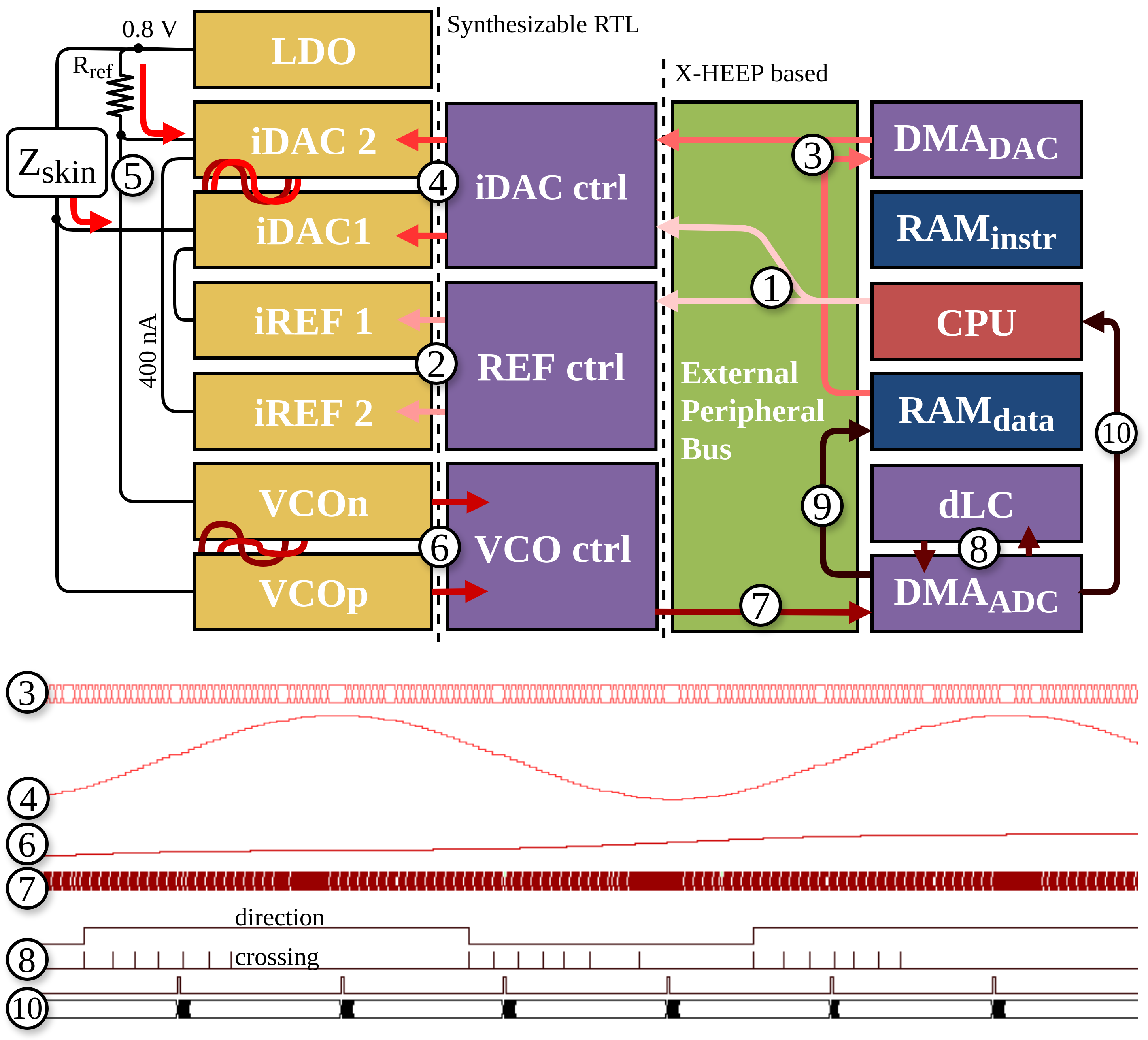}
\caption{Use case for impedance measurement showing the different stages and blocks involved (top). A behavioral simulation of the process (bottom) for a single VCO for simplicity.}
\label{fig:use-case}
\end{figure}

\begin{table*}[b!]
\centering
\caption{Comparison of HEEPidermis against SoA SoCs for BioZ recording}
\setlength{\tabcolsep}{3pt}
\renewcommand{\arraystretch}{1.25}
\begin{tabular}{>{\raggedright\arraybackslash}p{4cm} p{3cm} p{3cm} p{3cm} p{3cm}}
\hline\\[-7pt]
\textbf{Parameter} &
\makecell{\textbf{HEEPidermis}\\(this work)} &
\makecell{\textbf{MAX30001G}\\(Analog Devices)~\cite{MAX30001G}} &
\makecell{\textbf{ADS1292R}\\(Texas Instruments)~\cite{ADS1292R}} &
\makecell{\textbf{Rodriguez et al.}\\~\cite{rodriguez2015batteryless}}\\[5pt]
\hline\\[-7pt]
\makecell[l]{\textbf{Current injection}\\ (ch; freq; range; shape)} &
\makecell{2$\times$ DC–200\,kHz\\40\,nA–10\,$\mu$A\\arbitrary} &
\makecell{1$\times$ 125\,Hz–131\,kHz\\55\,nA–96\,$\mu$A\\square} &
\makecell{1$\times$ 32–64\,kHz\\30\,$\mu$A–100\,$\mu$A\\modulated} &
\makecell{1$\times$ 2\,kHz–2\,MHz\\3-bit @ 100mV\\ I/Q sine} \\[15pt]

\makecell[l]{\textbf{Digitization}\\(ch; bits; rate)} &
\makecell{2$\times$ 16-bit$^{*}$\\0.2\,mHz–10\,kHz} &
\makecell{1$\times$ 20-bit\\25–64\,Hz} &
\makecell{2$\times$ 24-bit\\125\,Hz–8\,kHz} &
\makecell{1$\times$ 10-bit\\I/Q demodulator} \\[10pt]

\textbf{Power} (acquisition) &
\makecell{72\,$\mu\text{W}^*$ @ 1.2/1.08\,V} &
\makecell{158\,$\mu$W @ 1.1\,V} &
\makecell{335\,$\mu$W/ch @ 3.0\,V} &
\makecell{300\,$\mu$W @ 1.8\,V} \\[5pt]

\textbf{On-chip processing} &
\makecell{32-bit RISC-V  CPU\\ event-based decimation} &
\makecell{Digital filtering\\decimation} &
\makecell{Digital filtering} &
\makecell{I/Q demod.}\\
\hline\\[-20pt]
\makecell{* Post-layout simulation results, impedance measurement case of Table I.}
\end{tabular}
\end{table*}

The impedance recording process is illustrated in \autoref{fig:use-case}: (1) The CPU sets the calibration of the references and iDACs and updates the frequency of the latter. (2) The references adopt the value. (3) The first DMA channel transfers 16-bit words containing a pair of de-phased sines to the iDACs --8-bit for each--. (4) Periodically, the iDACs adopt these values, controlled by an internal (configurable) timer. (5) The current is drained from the LDO through the tissue $\text{Z}_\text{skin}$ and the reference resistor $\text{R}_\text{ref}$. (6) The VCO-ADCs sense the signal. (7) The second DMA channel periodically acquires a single 32-bit output with the difference of the two channels. (8) The dLC compares the data against pre-determined thresholds, discards small variations, and sporadically outputs 8-bit words. (9) The dLC's output data is stored in memory. (10) After a fixed set of output values, the DMA wakes up the CPU to change the phase difference and re-launch the process, or compute the impedance. 

\begin{figure}[t!]
    \centering
    \includegraphics[width=0.8\linewidth]{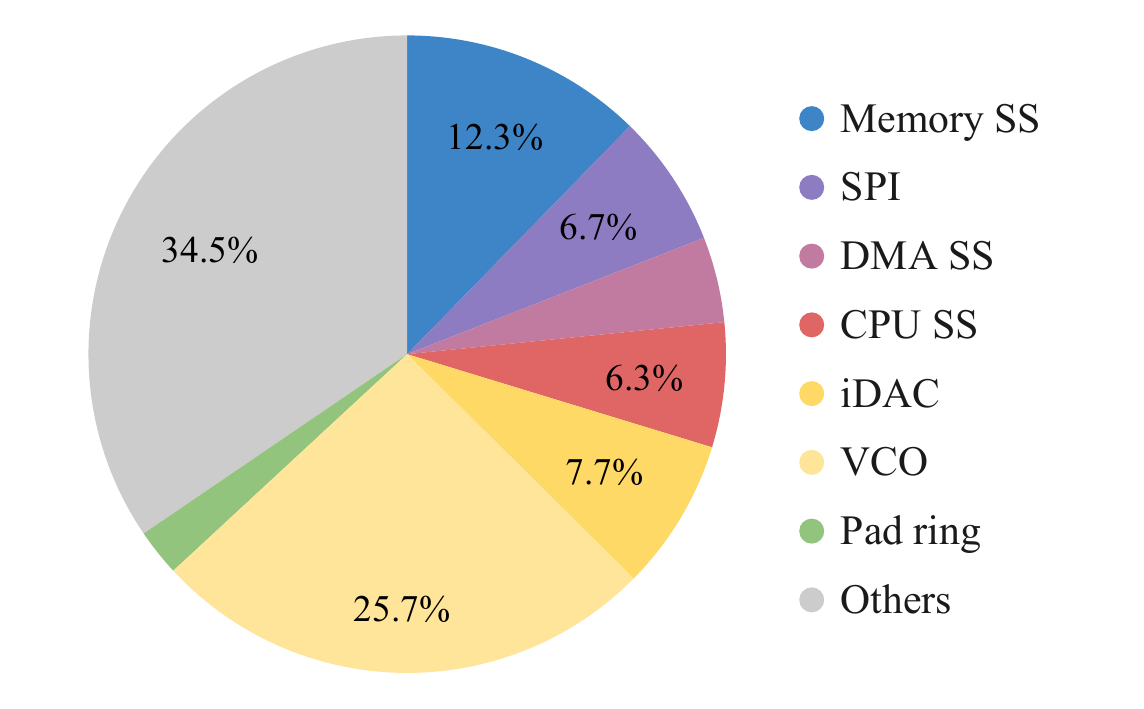}
    \caption{Power breakdown of the SoC during impedance measurement (total of \SI{72}{\micro\watt}).}
    \label{fig:power_bd}
\end{figure}

\autoref{tab:case1} summarizes the power breakdown for impedance recording (right) at a constant sensitivity of \SI{1}{\nano\siemens} in an average, intra-patient range of $10\,\mu-100\,\mu\text{S}$ (a dynamic range of 16 bits), using a \SI{5.12}{\micro\ampere} current, and sampling at \SI{20}{sps}. A finer breakdown for this case is shown in \autoref{fig:power_bd}. With the digital back-end running at \SI{1}{\mega\hertz} power is dominated by leakage (\SI{21}{\micro\watt}), and by non-clock-gated digital cells.

\section{Comparison with the State of the Art}

Table II compares HEEPidermis with other state-of-the-art SoCs for BioZ recording. Commercial SoCs are usually the building block for experimental setups implemented as custom PCBs~\cite{najafi2024versasens, AnalogDevices_MAXREFDES73_2016} unless they are implemented with off-the-shelf discrete blocks~\cite{canabal2020electrodermal,zhang2024advanced}. The main distinction from both commercial and research devices lies in their versatility: while existing solutions support a wide current range, they are constrained to fixed waveforms and limited frequency options, restricting experimental flexibility. In contrast, HEEPidermis supports arbitrary shapes and frequencies, broadening the range of possible studies. 
The inclusion of a programmable digital back-end allows long-term autonomous operation, including closed-loop interventions, without external hardware and at the lowest power budget. In addition, the system is open source, enabling further customization and specialization of the SoC to optimize performance for specific applications.

\section{Conclusion}

In this paper, we have presented HEEPidermis, an SoC to perform tissue impedance measurements. It integrates a versatile pair of iDACs that enable a wide range of research and therapeutic procedures. It also includes two ADC channels, allowing for a trade-off between power consumption for resolution and distortion for dynamic range. Using both in conjunction enables the adjustment of the operational point of the ADC. HEEPidermis also includes two DMA channels and an event-based sub-sampler block to free the CPU to wake up only when there are enough relevant data to process. %

\newpage
\printbibliography[]

\end{document}